\documentclass[twocolumn,aps,showpacs,eqsecnum,10pt]{revtex4} %twocolumn,
\usepackage{amsthm}
\usepackage{tikz}
\usetikzlibrary{arrows}
\usetikzlibrary{positioning}
\usetikzlibrary{calc}
\usepackage{amssymb}
\usepackage{graphicx}

\theoremstyle{definition}

\newcommand{\bra}[1]{\langle #1|}
\newcommand{\ket}[1]{| #1 \rangle }

\newcommand{\tr}[1]{{\rm tr}[#1]}
\newcommand{\eqref}[1]{{[\ref{#1}]}}

\newcommand{\be}{\begin{eqnarray}}
\newcommand{\ee}{\end{eqnarray}}

\newcommand{\norm}[1]{|| #1 ||}
\newcommand{\cE}{{\cal E}}

\newcommand{\cI}{{\cal I}}

\newcommand{\cF}{{\cal F}}

\newcommand{\cA}{{\cal A}}

\newcommand{\cL}{{\cal L}}

\def\arsi{5}
\def\minsi{6}
\def\dots{0.03}

\begin{document}
\title{Simulation of indivisible qubit channels in collision models}
\author{Tom\'a\v s Ryb\'ar$^1$, Sergey N. Filippov$^2$, M\'ario Ziman$^{3,1}$, Vladim\'\i r Bu\v zek$^1$}
\address{
$^{1}$Institute of Physics, Slovak Academy of Sciences, D\'ubravsk\'a cesta 9, 845 11 Bratislava, Slovakia \\
$^{2}$Moscow Institute of Physics and Technology, Moscow Region, Russia \\
$^{3}$Institute for Theoretical Physics, ETH Zurich, 8093 Zurich, Switzerland
}
%\ead{ziman@savba.sk}
\begin{abstract}
A sequence of controlled collisions between a quantum system and 
its environment (composed of a set of quantum objects) naturally simulates (with arbitrary precision) 
any Markovian  quantum dynamics of the system under consideration. In this paper we propose and study the problem 
of  simulation of an {\it arbitrary} quantum channel via collision models. We show
that a correlated environment is capable to simulate {\it non-Markovian} 
evolutions leading to any indivisible qubit channel. 
In particular, we derive the corresponding master 
equation generating a continuous time non-Markovian 
dynamics implementing the universal NOT gate being
an example of the most non-Markovian quantum channels.
\end{abstract}
%\submitto{\JPB}
\maketitle

%%%%%%%%%%%%%%%%%%%%%%%%%%%%%%%%%%%%%%%%%% introduction
\section{Introduction}
%%%%%%%%%%%%%%%%%%%%%%%%%%%%%%%%%%%%%%%%%%%%%%%%%%%%%%%%%%%%%%%%%%%%%%%%%
It is one of the basic postulates of quantum physics that
the time evolution of {\it closed} quantum systems is governed by
the Schr\"odinger equation \cite{schrodinger_equation}.
As a result for any time interval of length $\tau>0$
a state of a quantum system described by a density operator $\varrho$
is transformed according to a unitary conjugation
\be
\varrho \mapsto \varrho^\prime_\tau=U_\tau\varrho U^\dagger_\tau\,,
\ee
where $U_{\tau}=e^{-\frac{i}{\hbar}\tau H}$, $H$ is the (for simplicity
time-independent) Hamiltonian of the system and $\hbar$ is the Planck constant. 
Moreover, it is natural to assume that
$U_\tau=U_t U_s$, where $\tau=t+s$ and $t,s>0$ are 
arbitrary durations of short time intervals that sum up to 
the whole time interval $\tau$. However, if the system is open (interacting with
its environment) such simple division of time evolution into incremental time steps does not necessarily 
possess a clear meaning. It is a relatively recent discovery
\cite{wolf_indivisible} that certain non-unitary quantum evolutions 
are indivisible into shorter ones. In order to make this statement precise 
let us introduce a mathematical framework modelling open system dynamics 
over finite time intervals. 

Let us assume that initially the system and its environment 
are not correlated, so that that the combined system-environment 
initial state can be described as $\varrho\otimes\xi $
(such an assumption is quintessential in order to
preserve linearity of the dynamics - see eg Ref. 
\cite{stelmachovic_buzek} and references therein). 
Then the evolution of the joint system-environment state is governed 
by the Schr\"odinger equation and the state transformation over 
the time interval $[0,\tau]$ is described by a map
\be
\varrho\mapsto\cE_\tau[\varrho]={\rm tr}_{\rm env}[\tilde{U}_\tau\varrho\otimes\xi \tilde{U}_\tau^\dagger]\,,
\ee 
where $\xi$ is the initial state of the environment and 
${\rm tr}_{\rm env}$ denotes the partial trace over the environmental
degrees of freedom and $\tilde{U}_\tau$ is a unitary transformation describing 
both the system and the environment.  
Abstracted mathematical properties of $\cE_\tau$ 
guaranteeing the existence of such model for $\cE_\tau$ are 
linearity ($\cE(X+\lambda Y)=\cE(X)+\lambda\cE(Y)$), 
complete positivity (positivity of $\cE_\tau\otimes\cI_d$ for all $d$) and 
trace-preservation ($\tr{\cE_\tau[X]}=\tr{X}$). 
This result is known as the  \emph{Stinespring representation theorem} 
\cite{stinespring} and the maps $\cE_\tau$ with such properties we call 
\emph{quantum channels} (see for example \cite{heinosaari_qbook}).

We say that a quantum channel $\cE$ is divisible if $\cE=\cE_1\circ\cE_2$,
where $\cE_1,\cE_2$ are both non-unitary quantum channels. If this
is not the case, then we say the channel is indivisible. 
According to this definition unitary channels are indivisible. This
may sound quite counterintuitive, especially,  if we recall how
unitary operations are decomposed into gates in quantum computation \cite{nielsen2000}.
Indeed the gate  decomposition
is the crucial tool in the analysis of quantum complexity. 
But let us stress a difference between the decomposition and the divisibility 
of channels. The primary task of the decomposition is to simplify the 
implementation of a multipartite quantum channel by means of 
elementary gates, thus, the decomposability captures 
the complexity of the process expressed in terms of number of uses of elementary gates. 
On the other side, the divisibility 
addresses more fundamental question whether a given process can be understood
as a sequential concatenation of processes. If not, then it is natural
to ask how to simulate such quantum channels, in particular by means
of some "continuous" time evolution.

Dynamics of open quantum systems is modelled by so-called master
equations that are usually derived under the assumption of Markovianity
\cite{davies_book,breuer_book}.
Mathematically this means that one parametric set of channels 
$\cE_t$, being the solution of Markovian master equation (equivalently 
Lindblad master equation), satisfies
the semigroup property $\cE_{s}\circ\cE_t=\cE_{s+t}$ for all $s,t\geq 0$
and the initialization condition $\lim_{t\to 0}\cE_{t}=\cI$. Formally
we can write $\cE_t=e^{\cL t}$, where $\cL$ is known as the Lindbladian.
We say a channel $\cE$ is Markovian if there exist the Lindbladian $\cL$
and time $\tau$ such that $\cE=e^{\cL\tau}$. Otherwise, the channel
is called non-Markovian. Let us note that all indivisible channels
are non-Markovian, but there are divisible non-Markovian channels 
\cite{wolf_indivisible,wolf_snapshot}.

Simple collision models, in which the system's evolution 
is modelled via a sequence of (weak) interactions of the system
with sequentially selected particles from the environment,  
provide a natural playground for simulations of quantum dynamics.
It is known (see for instance Refs. \cite{ziman_collision,ziman_chapter}) that 
these models can approximate (with arbitrary precision) 
any evolution governed by a Lindblad master equation \cite{lindblad,gorini}, 
i.e. implement any Markovian quantum channel. Our aim in this paper
is to investigate whether a general quantum channel (in particular
an indivisible one) can be simulated in the framework 
of simple collision models. Let us note that a related
collision model was studied in Ref. \cite{giovannetti2012}, where
the authors derived master equations for a family of correlated Markovian 
evolutions.

This paper is structured as follows. In Section II, we define the
framework of collision models and we introduce the concept
of stroboscopic simulations. In Section III, we design 
a collision model for implementation of {\it all} indivisible 
qubit channels. Finally, we discuss our results and present conclusions of our investigation 
in Section IV.

%%%%%%%%%%%%%%%%%%%%%%%%%%%%%%%%%%%%%%%%%%%%%%%%%%%%%%%%%%%%%%%%%%%%%%%%% 
\section{Simple collision model}
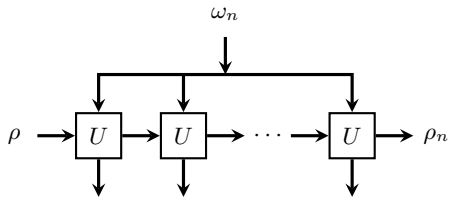
\begin{figure}[h]
\begin{center}
\begin{tikzpicture}[
	box/.style={
		% The shape: 
		rectangle,
		% The size: 
		minimum size=\minsi mm,
		% The border:
			thick, draw=black,		 
		node distance=\arsi mm,
		% Font 
		font=\itshape
	},
	unbox/.style={
		% The shape: 
		rectangle,
		% The size: 
		minimum size=\minsi mm,
		% The border:		 
		node distance=\arsi mm,
		% Font 
		font=\itshape
	},
	dot/.style={
		% The shape: 
		rectangle,
		% The size: 
		minimum size=\minsi mm,
		% The border: 
		node distance=\arsi mm,
		% Font 
		font=\itshape
	},
	arr/.style={
		very thick, draw=black,		 
	},
	>=stealth
	]
	
	\node[unbox] (memin) {$\rho$};
	\node[box] (1) [right=of memin] {$U$};
	\node[box] (2) [right=of 1] {$U$};
	\node[dot](3) [right=of 2] {$\cdots$};
	\node[box](6) [right=of 3] {$U$};
	\node[unbox](7) [right=of 6] {$\rho_n$};
	\draw[arr,->] (memin.east)--(1.west);
	\draw[arr,->] (1.east)--(2.west);
	\draw[arr,->] (2.east)--(3.west);
	\draw[arr,->] (3.east)--(6.west);
	\draw[arr,->] (6.east)--(7.west);
	\draw[arr,<-](2.north)--($(2.north)+(0 mm,\arsi mm)$);
	\draw[arr,->](1.south)--($(1.south)-(0 mm,\arsi mm)$);
	\draw[arr,->](2.south)--($(2.south)-(0 mm,\arsi mm)$);
	\draw[arr,->](6.south)--($(6.south)-(0 mm,\arsi mm)$);
	\coordinate (l) at ($(6.north)+(0 mm,\arsi mm)$);
	\coordinate (r) at ($(1.north)+(0 mm,\arsi mm)$);
	\draw[arr,<->](1.north)-- (r)--(l)--(6.north);
	\coordinate (m) at ($0.5*(l)+0.5*(r)$);
	\node[unbox] (8) [above=of m] {$\omega_n$};
	\draw[arr,->] (8.south)--(m);
	\end{tikzpicture}
\caption{Schematic illustration of a simple collision model.
\label{collision_model}}
\end{center}
\end{figure}

%%%%%%%%%%%%%%%%%%%%%%%%%%%%%%%%%%%%%%%%%%%%%%%%%%%%%%%%%%%%%%%%%%%%%%%%% 
By a collision model of a specific system dynamics we understand
a sequence of interactions (collisions) between the system and
particles from the environment. Essential property of the model considered
in this paper is that each particle of the environment interacts with the system at most
once, while the environment particles do not interact between themselves. 
More formally, let
$U_j$ be a bi-partite unitary operator describing an individual collision between the system particle and the $j$-th particle of the environment. 
Collision model is a concatenation of unitary channels $U_1\dots U_n$ (see Fig.~\ref{collision_model}).
Let us denote by $\omega_n$ the initial state of the environment
composed of $n$ particles and by $\varrho$ the initial state
of the system.

Suppose $\varrho\to\varrho_t=\cE_t(\varrho)$ is a solution of 
some Lindblad master equation. The question we will like to answer is ``How to simulate
the process $\cE_t$ in the framework of a collision model?'' Assume $U_{\delta}$ 
and $\xi$ determine the Stinespring dilation of $\cE_{\delta}$. 
Then the sequence of concatenations of $U_\delta$ applied in each 
step jointly on the system and a ``fresh'' particle from the environment 
in the state $\xi$,  results in a discrete evolution
\be
&&\varrho\mapsto\varrho_n=\cE_{\delta}^n(\varrho)\nonumber\\
&&=
{\rm tr}_{\rm env}[(U_1\cdots U_n)(\varrho
\otimes\xi^{\otimes n}) (U_1\cdots U_n)^\dagger]\,,
\ee
where $U_j=U_\delta\otimes I_{\overline{j}}$ and $I_{\overline{j}}$
denote the identity operator on all particles of the environment
but the $j$th one. As a result the described 
collision model \emph{stroboscopically simulates} the continuous Markovian 
evolution, i.e. for all $\delta>0$ we have $\cE_t\approx\cE_{\delta}^n$ 
with $n=[t/\delta]$ ($[x]$ denotes the nearest integer to $x$). 
The parameter $\delta$ determines the quality of the (stroboscopic) 
approximation of the time-continuous (Markovian) dynamics $\cE_t$ 
for $t\geq 0$. 

It was shown in \cite{ziman_deco} that if $U$ is a controlled unitary
operator, then the sequence of collisions simulates pure decoherence process,
in which the diagonal elements of density operators 
(with respect to the so-called decoherence basis) are preserved
while the off-diagonal ones vanish exponentially. Setting $U$
to be the partial SWAP interaction \cite{ziman_homo,scarani_thermo} the system 
exponentially converges to the original state of the particles 
in the reservoir. The information about the original state of the 
system is diluted into the correlations among environmental particles.
Because of these features the process is called quantum homogenization and
represents a quantum information analogue of thermalization process, in which
the temperature is replaced by the concept of quantum state. 

In both examples described above it is assumed that the initial state of the 
reservoir is factorized, i.e. $\omega_n=\xi^{\otimes n}$. In this factorized 
setting, the channel $\cE_1$ is called a \emph{generating channel} 
of the collision process. Indeed, if this is
the case, then the 
induced discrete dynamics $\cE_n\equiv\cE_1^n$ fulfills the conditions 
of a discrete semigroup,
i.e. $\cE_n\circ\cE_m=\cE_{n+m}$ for all positive integers $n,m$. Without
loss of generality we may set $\cE_0=\cI$.  It is important
to stress that such semigroup feature does not guarantee that 
the discrete dynamics stroboscopically approximates some
Markovian continuous-time evolution. In other words, the generated
channels $\cE_n$ are not necessarily Markovian.
For instance, by definition indivisible channels $\cE_{\rm indivisible}$ can 
be elements of a discrete semigroup only as the generating channels, i.e.
$\cE_{\rm indivisible}=\cE_1$. 

But what if a general
initial state of the environment is allowed? Could we then find 
a way how to stroboscopically simulate (using the collision model) 
a continuous-time evolution leading to an arbitrary quantum channel?

Let us formulate the problem: We say $\cE$ is a 
\emph{stroboscopically simulated channel} if for every $n$ 
there exists a bipartite interaction $U$ 
(between the system and a specific particle from the environment ) 
and an initial state $\omega_n$ of the environment 
such that $\cE_n\equiv\cE$, where
$$
\cE_n(\varrho)=
{\rm tr}_{\rm env}[(U_1\cdots U_n)(\varrho\otimes\omega_n)
(U_1\cdots U_n)^\dagger]\,,
$$
and $U_j=U\otimes I_{\overline{j}}$. If the channel can be 
approximated in the introduced sense, then
the collision models reveal the elementary features of Markovian
dynamics even for non-Markovian channels. In particular,
the process can be realized in arbitrarily small (non-unitary) 
steps although the steps themselves are not described by valid
channels. Our aim is to address the question which channels 
can be stroboscopically simulated? In what follows we will focus
on implementation of indivisible quantum channels, which are, intuitively,
the most non-Markovian examples of quantum evolutions.

%%%%%%%%%%%%%%%%%%%%%%%%%%%%%%%%%%%%%%%%%%%%%%%%%%%%%%%%%%%%%%%%%%%
\section{Stroboscopic simulation of indivisible qubit channels}
%%%%%%%%%%%%%%%%%%%%%%%%%%%%%%%%%%%%%%%%%%%%%%%%%%%%%%%%%%%%%%%%%%%
The existence of indivisible channels was observed for the 
first time in Ref.~\cite{wolf_indivisible}, where the authors
analyzed in detail qubit channels. In particular,
they found that indivisible qubit channels are unitarily equivalent 
to the subclass of Pauli channels
\be
\label{eq:indivisible_qubit_family}
\cE(\varrho)=q_x\sigma_x\varrho\sigma_x+
q_y\sigma_y\varrho\sigma_y+q_z\sigma_z\varrho\sigma_z\,
\ee 
where $q_x,q_y,q_z$ are positive ($q_xq_yq_z\neq 0$) 
and $q_x+q_y+q_z=1$. For instance, the choice $q_x=q_y=q_z=1/3$ 
defines the best quantum approximation $\cE_{\rm NOT}$
of the universal quantum NOT gate for a qubit $\cF_{\rm NOT}:\varrho\mapsto 
(I-\varrho)$ \cite{buzek_not, martini}. The question is: how to simulate 
stroboscopically some continuous time dynamics leading to these channels?

Suppose environment is composed of three-dimensional quantum systems
(qutrits). Then one of the possible dilations for indivisible qubit 
channels is the following:
$\cE(\varrho)=
{\rm tr}_{\rm env}[C\varrho\otimes\omega C^\dagger]$, where
$C=\sigma_x\otimes\ket{x}\bra{x}+\sigma_y\otimes\ket{y}\bra{y}+
\sigma_z\otimes\ket{z}\bra{z}$ is a controlled unitary interaction
and $\omega$ satisfies the conditions $\bra{k}\omega\ket{k}=q_k$
for $k=x,y,z$. We assume that $\ket{x},\ket{y},\ket{z}$ form
an orthonormal basis of the qutrit's Hilbert space. Let us stress 
that $C$ is not only unitary, but also Hermitian. Therefore
$U_\eta=e^{i\eta C}=\cos\eta I+i\sin\eta C$ defines a family 
of interactions. Moreover, each member of this family is 
a controlled-U operator, $U_\eta=\sum_k e^{i\eta \sigma_k}
\otimes\ket{k}\bra{k}$.

Consider a collision model generated by collisions $U_\eta$. 
The concatenation of $n$ such collisions
implements a global (system plus environment) unitary transformation 
\be
U_{\eta,1}\cdots U_{\eta,n}=\sum_k \exp[i n\eta \sigma_k]
\otimes\ket{k^{\otimes n}}\bra{k^{\otimes n}}\,.
\ee
Suppose $\omega_n$ is the initial state of the environment composed of $n$ qutrits  
such that $\bra{k^{\otimes n}}\omega_n\ket{k^{\otimes n}}=q_k$.
After $n$th collision the system evolution reads
\be
\nonumber
\cE_n(\varrho)&=&
{\rm tr}_{\rm env} [(U_{\eta,1}\cdots U_{\eta,n})(\varrho\otimes\omega_n) 
(U_{\eta,1}\cdots U_{\eta,n})^\dagger]\\
\nonumber &=&\sum_k q_k \exp[in\eta\sigma_k]\,\varrho\, \exp[-in\eta\sigma_x]\,,
\ee
hence setting the strength of the interaction $\eta=\pi/(2n)$ the
resulting collision model simulates the implementation of 
any indivisible qubit channel in $n$ steps.

\subsection{Evaluation of $\norm{\cE_{j+1}-\cE_j}$}
In this Section we will show that for the considered
collision model the individual collisions
induce arbitrarily small disturbances of the system. 
Assume $n$ is fixed, i.e. $\eta=\pi/(2n)$. 
After $j$th interaction the state of the system undergoes
the transformation
\be
\nonumber
\cE_j(\varrho)&=&\sum_k q_k \exp[i\pi\frac{j}{2n}\sigma_k]\,\varrho\, 
\exp[-i\pi\frac{j}{2n}\sigma_k]\\
\nonumber
&=&\cos^2(\frac{j\pi}{2n}) \varrho+
\sin^2(\frac{j\pi}{2n})\sum_k q_k\sigma_k\varrho\sigma_k
\nonumber\\
&&+\frac{i}{2}\sin(\frac{j\pi}{n})\sum_k q_k[\sigma_k,\varrho]\,,
\ee
where $[A,B]=AB-BA$ is the commutator of operators $A,B$.
Let us denote by $\cE$ the target quantum channel, i.e.
$\cE=\cE_n$ (we fix $\eta=\pi/(2n)$) and
define $\cF(\cdot)=[F,\cdot]$ with $F=i\sum_k q_k\sigma_k$. 
Then the distance between two subsequent steps equals
\be
\Delta&=&\norm{\cE_{j+1}-\cE_j}=
\norm{
C_{j+1}(\cI-\cE)+D_{j+1}\cF}\,,
\ee
where
\be
C_{j+1}&=&\cos^2\frac{(j+1)\pi}{2n}-\cos^2\frac{j\pi}{2n}\nonumber\\
&=&\frac{1}{2}\left(\cos\frac{(j+1)\pi}{n}-\cos\frac{j\pi}{n}\right)\,,\\
D_{j+1}&=&\frac{1}{2}\left(\sin\frac{(j+1)\pi}{n}-\sin\frac{j\pi}{n}\right)\,.
\ee
Let us note that the norm we have in mind here is the completely
bounded norm (see for instance \cite{paulsen2003}), namely,
$$
\norm{\cA}=\sup_{n,X:||X||_1=1}||(\cA\otimes\cI_n)(X)||_1\,,
$$
where $||\cdot||_1={\rm tr}|\,\cdot\,|$.

Using the identities $\cos^2\alpha=(1+\cos{2\alpha})/2$,
$\cos\alpha-\cos\beta=-2\sin(\frac{\alpha+\beta}{2})
\sin(\frac{\alpha-\beta}{2})$, and
$\sin\alpha-\sin\beta=2\sin(\frac{\alpha-\beta}{2})
\cos(\frac{\alpha+\beta}{2})$
we get
\be
C_{j+1}&=&-\sin\frac{(2j+1)\pi}{n}\sin\frac{\pi}{n}\,,\\
D_{j+1}&=&\cos\frac{(2j+1)\pi}{n}\sin\frac{\pi}{n}\,.
\ee
This allows us to conclude that the distance is bounded as follows:
\be
\nonumber
\Delta&\leq&
|C_{j+1}|\cdot\norm{\cI-\cE}+|D_{j+1}|\cdot\norm{\cF}\\
\nonumber
&\leq& \sin\frac{\pi}{n}\left(
\left|\sin\frac{(2j+1)\pi}{n}\right| \cdot\norm{\cI-\cE}+\left|\cos\frac{(2j+1)\pi}{n}\right|\cdot\norm{\cF}\right)\\
&\leq& K\sin\frac{\pi}{n}\quad(\,\longrightarrow 0\quad {\rm for\ }n\to\infty)\,,
\ee
where $K\leq\norm{\cI-\cE}+\norm{\cF}\leq 2+\norm{\cF}<\infty$,
because for channels $\norm{\cE}=1$ and $\norm{\cF}<\infty$. 
In fact, as shown in Ref. \cite{paulsen2003} for any linear qubit map 
$\norm{\cA}\leq 2\sqrt{2} \sup_{\norm{X}_1=1} \norm{\cA(X)}_1$, hence,
$\norm{F}\leq 2\sqrt{2} \sup_{\norm{X}_1=1}\norm{FX-XF}_1\leq
4\sqrt{2}\frac{\norm{FX}_1}{\norm{X}_1}\leq 4\sqrt{2}\norm{F}$,
where the relations $\norm{FX}_1\leq \norm{F}\cdot\norm{X}_1$
and $\norm{F}=\sup_\psi \norm{F\psi}/\norm{\psi}\leq 2$ were used. That is,
$K\leq 2+8\sqrt{2}$, which allows us to
conclude that the system's changes in individual steps 
can be made arbitrarily small as $n$ goes to infinity.
Therefore, in this limit the evolution is continuous.

\subsection{Master equation}
Replacing the integer parameter $j$ in the expression for $\cE_j$
by a continuous parameter $t$ we formally define a one-parametric 
set of channels 
$$
\cE_t=\cE+\cos^2(\alpha t)(\cI-\cE)+\sin(\alpha t)\cos(\alpha t)\cF\,,
$$
where $\alpha=\pi/(2n)$ and linear maps $\cE,\cF$ are defined in the
previous paragraph. Let us note that $\cE_t$ form the same set of channels
irrelevant of the value of $\alpha$. Moreover, this one-parametric set
of channels is continuous in $t$ and $\cE_0=\cI$. A collision model 
(determined by the value of $\alpha$) stroboscopically
simulates continuous-time quantum evolution given by $\cE_t$, where
different values of $\alpha$ define the quality of the simulation.
In this section we will derive the master equation generating $\cE_t$.

Formally, the evolution of density operators is generated by 
the first order differential equation
\be
\frac{d\varrho_t}{dt}=\frac{d\cE_t}{dt}\cE_t^{-1}(\varrho_t)\equiv
\cL_t(\varrho_t)\,,
\ee
where
\be
\cL_t(X)&=&
\frac{i}{\hbar}\sum_j h_j [X,\sigma_j]+\frac{1}{2}
\sum_{j,k}c_{jk}([\sigma_j,X\sigma_k]\nonumber\\
&&+[\sigma_jX,\sigma_k])
\ee
is the generator of the dynamics \cite{lindblad,gorini}. 
Any such generator defines
a dynamics which is tracepreserving and if (time-dependent coefficients)
$h_j,c_{jk}$ are real, then also the hermiticity of operators is
preserved. If these parameters are time-independent ($\cL_t=\cL$)
and matrix composed of entries $c_{jk}$ is positive, then the generated 
dynamics is also completely positive and Markovian. In such case
we can write $\cE_t=e^{\cL t}$ and $\cL$ is the Lindbladian.

For sake of simplicity let us illustrate the
derivation of the driving master equation ($\cL_t$) for the case of 
the target channel $\cE=\cE_{\rm NOT}$. This channel transforms 
the Bloch vector $\vec{r}$ into $\vec{r}^\prime=-\vec{r}/3$, hence, 
implements a Bloch ball (shrinking) inversion. In this case the map 
$\cF$ induces the Bloch vector transformation 
$\vec{r}\to\vec{u}\times\vec{r}$ with $\vec{u}=(1/3,1/3,1/3)$.
Let us stress that $\cF(I)=O$, hence, the map is not 
trace-preserving and maps any operator into a traceless one.

\begin{figure}[h]
\begin{center}
\includegraphics[width=8cm]{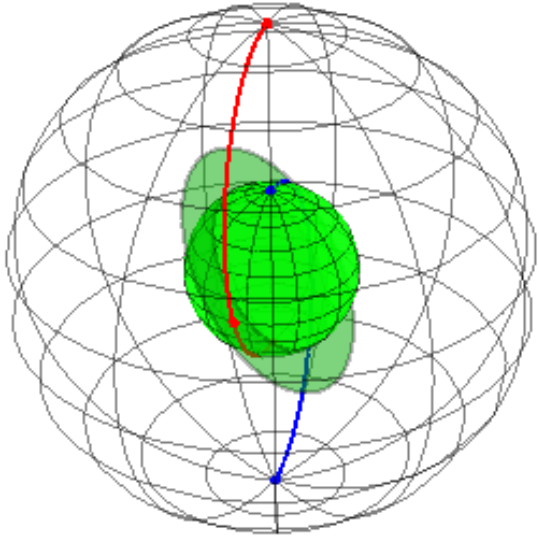}
\caption{The collision model simulating the continuous 
time evolution towards the universal NOT gate (shrinked Bloch sphere 
inversion). In particular, the transformation of the Bloch sphere 
(lines capture the time evolution of eigenstates of $\sigma_z$ operator) 
is depicted for the time interval $t\in[0,n]$. For $t=\frac{2}{3}n$ 
the channel $\cE_t$ in not invertible ($\det{\cE_{t=2n/3}}=0$) 
and at this time the Bloch sphere is mapped onto a two-dimensional 
disk. Let us note that images of eigenstates of $\sigma_z$ operator
are internal points of the disk. In fact, the whole disk is the 
image of pure states only.
\label{fig:NOT_collision}
}
\end{center}
\end{figure}

In the Bloch sphere parametrization the channels take the form
of an affine $4\times 4$ matrix. Define $3\times 3$ matrices
\be
I&=&\left(
\begin{array}{ccc}
1 & 0 & 0\\
0 & 1 & 0\\
0 & 0 & 1
\end{array}\right)\,,\quad
A=\left(
\begin{array}{ccc}
0 & -1 & 1\\
1 & 0 & -1\\
-1 & 1 & 0
\end{array}\right)\,,
\nonumber\\
S&=&\left(
\begin{array}{ccc}
0 & 1 & 1\\
1 & 0 & 1\\
1 & 1 & 0
\end{array}\right)\,.
\ee
Then the one-parametric dynamics is given by the following
$4\times 4$ matrix
$$
\cE_t=
\left(\begin{array}{cc}
1 & 0 \qquad 0 \qquad 0 \\
\vec{0} & x(t)I+a(t)A
\end{array}\right)\,,
$$
where $x(t)=\frac{1}{3}(4\cos^2\alpha t -1)$
and $a(t)=\frac{1}{6}\sin(2\alpha t)$.
A direct calculation gives
\be
\frac{d\cE_t}{dt}=\frac{1}{3}\alpha
\left(\begin{array}{cc}
0 & 0\qquad 0 \qquad 0 \\
\vec{0} & 4\sin (2\alpha t) I+\cos (2\alpha t) A \\
\end{array}\right)\,,
\ee
and
\be
\cE_t^{-1}=
\left(\begin{array}{cc}
1 & 0\qquad 0 \qquad 0 \\
\vec{0} & \frac{1}{3x(t)}I-\frac{a(t)}{\det{\cE_t}}
[x(t)A-a(t)S] \\
\end{array}\right)\,,
\ee
where $\det{\cE_t}=3x(t)[a(t)^2+x(t)^2]$. Thus, for the generator
we get
\be
\cL_t=\frac{d\cE_t}{dt}\cE_t^{-1}=
\left(\begin{array}{cc}
0 & 0\qquad 0 \qquad 0 \\
\vec{0} & b(t)I+c(t)A+d(t)S \\
\end{array}\right)\,,
\ee
with
\be
b(t)&=& \frac{2}{9}a(t)\left[12\alpha+\frac{1}{a(t)^2+x(t)^2}\right]\,,\\
c(t)&=& \frac{1}{9x(t)}\left[\frac{\alpha (3x(t)-1)}{2}-
\frac{a(t)[3x(t)+a(t)]}{x(t)[a(t)^2+x(t)^2]}\right]\,,\nonumber\\
\\
d(t)&=& \frac{a(t)[3a(t)-x(t)]}{9x(t)[a(t)^2+x(t)^2]}\,.
\ee
Using the methods and formulas described in Ref.~\cite{ziman_chapter}
we obtain the non-Markovian master equation in the operator form
\be
\frac{d\varrho_t}{dt}&=&-\frac{ic(t)}{2\hbar}[\varrho_t,H]
-\frac{b(t)}{2}(\sum_j \sigma_j\varrho_t\sigma_j - 3\varrho_t)\nonumber\\
&&+d(t)\sum_{j\neq k}{\sigma_j\varrho_t\sigma_k}\,,
\ee
where $H=\sigma_x+\sigma_y+\sigma_z$. 
The time dynamics induced by the collision model is illustrated 
in Fig.~\ref{fig:NOT_collision}. 

%%%%%%%%%%%%%%%%%%%%%%%%%%%%%%%%%%%%%%%%%%%%%%%%%%%%%%%%%%%%%%%%%%%
\section{Conclusions}
%%%%%%%%%%%%%%%%%%%%%%%%%%%%%%%%%%%%%%%%%%%%%%%%%%%%%%%%%%%%%%%%%%%
In this paper we open 
the question of continuous time (stroboscopic) simulation of 
quantum channels within simple collision models, i.e. via a sequence 
of interactions of the system with particles forming the environment. 
Using the environment composed of three-dimensional quantum
particles we design a collision model simulating arbitrary indivisible 
qubit channel. Indivisible channels could be coined as the most 
non-Markovian ones if one quantifies the \emph{Markovianity of a channel} 
$\cE$ as the maximal number ($n$) of non-unitary channels 
$\cE_1,\dots,\cE_n$ such that $\cE=\cE_1\cdots\cE_n$. The smaller the 
number, the more non-Markovian is the channel. Let us stress that
there are several recent proposals for measures of Markovianity 
of continuous time evolutions 
\cite{breuer2009_nonmarkovianity_measure,rivas2010,{usha_devi_2011}}. 
These measures can be applied not only to the derived master equation for
implementation of $\cE_{\rm NOT}$, but can be also 
modified to the settings of discrete time evolutions, which are 
naturally generated by simple collision models. However, a more 
detailed analysis along these lines goes beyond the scope 
of this paper.

Another important point we want to stress is the necessity
of initial correlations between the particles of the 
reservoir, which are introducing the memory mechanism present in
any non-Markovian evolution. It is of interest to understand
whether there is some deeper relation between the correlation
structure of $\omega_n$ and memory features of the system's 
dynamics (Markovianity). In order to answer this question 
one needs to understand the ambiguity of the stroboscopic simulation. 
For example, each convex decomposition into unitary channels
induces a different collision model (as described in the previous Section) 
for the same channel $\cE$. However, for this class of collision models
there is no qualitative difference neither in the initial states, nor
in the derived evolutions. In the considered collision model, 
the correlations are relatively 
strong and all particles forming the environment are mutually pairwise
correlated, but it could happen that there are qualitatively different 
collision models for which the structure of correlations is completely 
different, especially, it could be that particles entering the $j$th 
and the $k$th collision are initially uncorrelated if $|j-k|$ is 
sufficiently large. 

In summary, we have introduced the problem of stroboscopic simulations of a general
quantum channels. In this paper we reported the case study of qubit
indivisible channels and we designed a collision model for implementation
of any channel from the family of random unitary channels
(see \ref{A}). Since indivisible channels are the most non-Markovian,
it is natural to conjecture that any channel can be simulated 
in the stroboscopic manner. In fact, it was shown in 
\cite{wolf_indivisible} that 
nonunital qubit channels are infinitesimal divisible and 
can be approximated by a concatenation of Markovian 
channels that can be stroboscopically simulated by 
factorized states of the environment. Moreover, since unital qubit channels are necessarily 
random unitary and  for them we have an explicit collision model 
(irrelevant of their Markovianity), we may conclude that collision 
models can stroboscopically approximate any qubit channel. 
We believe the considered collision model 
deserves further investigation that finally results in a better understanding 
of non-Markovian features of general continuous time quantum evolutions
and implementations of non-Markovian quantum channels.

%%%%%%%%%%%%%%%%%%%%%%%%%%%%%%%%%%%%%%%%%%%%%%%%%%%%%%%%%%%%%%%%%%%%%%%%% 
\section*{Acknowledgments}
We acknowledge financial support of the European Union project 
2010-248095 (Q-ESSENCE), COST action MP1006 and CE SAS QUTE.
T.R. acknowledges the support of APVV LPP-0264-07 (QWOSSI). 
S.N.F. thanks the Russian Foundation for Basic Research 
(projects 10-02-00312 and 11-02-00456), the Dynasty
Foundation, and the Ministry of Education and Science of the
Russian Federation (projects 2.1.1/5909, $\Pi$558, 2.1759.2011 and
14.740.11.1257). 
M.Z. acknowledges the support of SCIEX Fellowship 10.271.

%%%%%%%%%%%%%%%%%%%%%%%%%%%%%%%%%%%%%%%%%%%%%%%%%%%%%%%%%%%%%%%%%%%%%%%%%

\appendix
\section{Simulation of all random unitary channels}
\label{A}
In this Appendix we will extend the presented stroboscopic simulation
of qubit indivisible channels to any random unitary channel acting on the system of arbitrary dimension. 
A channel is called random unitary 
if $\cE=\sum_j q_j V_j\varrho V_j^\dagger$, $q_j\geq 0$, $\sum_j q_j=1$ and
$V_j$ are unitary operators. Suppose $1\leq j \leq d$ and define a collision
model generated by the interaction
$$
U=\sum_{j=1}^d\ket{j}\bra{j}\otimes {V_j}^{1/n}\,,
$$
where $V_j^{1/n}$ are unitary operators such that 
$(V_j^{1/n})^n=V_j$. Assuming that initially
$\omega_j=\sum_j q_j \ket{j^{\otimes n}}\bra{j^{\otimes n}}$,
we find (analogously as 
for the qubit case) that after $k$th collision
\be
\cE_k(\varrho)=\sum_{j=1}^d q_j\, (V_j^{k/n})\,\varrho\, (V_j^{k/n})^{\dagger}\,,
\ee
thus, $\cE_n=\cE$. Such construction works for any value of $n$
and therefore we can conclude that arbitrary random unitary channel
can be stroboscopically simulated. Let us note that this
collision model defines a non-Markovian evolution also for 
Markovian channels.


\section*{References}
\begin{thebibliography}{99}

\bibitem{schrodinger_equation}
E.Schr\"odinger,
%An Undulatory Theory of the Mechanics of Atoms and Molecules,
Phys. Rev. 28, 1049–1070 (1926). 

\bibitem{wolf_indivisible}
M.M. Wolf, and J.I. Cirac,
%Dividing quantum channels, 
Commun. Math. Phys. 279, 147 (2008). 
%[arXiv:math-ph/0611057]

\bibitem{stelmachovic_buzek}
P. \v{S}telmachovi\v{c} and V. Bu\v{z}ek, 
%Dynamics of open quantum systems initially entangled with environment: Beyond the Kraus representation 
Phys. Rev. A 64, 062106 (2001)

\bibitem{stinespring}
W.F. Stinespring, 
%Positive Functions on C*-algebras, 
Proceedings of the American Mathematical Society, 211–216 (1955).

\bibitem{heinosaari_qbook}
T. Heinosaari, and M. Ziman,
\emph{The Mathematical Language of Quantum Theory},
(Cambridge University Press, Cambridge, 2012).

\bibitem{nielsen2000}
M.A. Nielsen and I.L. Chuang,
{\it Quantum Computation and Quantum Information},
(Cambridge University Press, Cambridge, 2000).

\bibitem{davies_book}
E.B.Davies, 
\emph{Quantum Theory of Open Systems},
(Academic Press Inc, London 1976)

\bibitem{breuer_book}
H.-P. Breuer, and F.Petruccione,
\emph{The Theory of Open Quantum Systems},
(Oxford University Press, 2002)

\bibitem{wolf_snapshot}
M.M. Wolf, J. Eisert, T.S. Cubitt, and J.I. Cirac,
%Assessing non-Markovian dynamics
Phys. Rev. Lett. 101, 150402 (2008).  
%[arXiv:0711.3172].

\bibitem{ziman_collision}
M. Ziman, P. \v Stelmachovi\v c, and V.Bu\v zek,
%Description of quantum dynamics of open systems based on collision-like models,
Open systems and information dynamics 12, 81-91 (2005).
%[quant-ph/0410161].

\bibitem{ziman_chapter}
M.Ziman, and V.Bu\v zek,
%Open system dynamics of simple collision models
chapter in 
\emph{Quantum Dynamics and Information (Proceedings of 46th Karpacz Winter School of Theoretical Physics)}, 
(World Scientific Publishing, Singapore, 2011).
%[arXiv:1006.2794].

\bibitem{lindblad}
G. Lindbald, 
Commun. Math. Phys. 48, 119 (1976).

\bibitem{gorini}
V. Gorini, A. Kossakowski, and E.C.G. Sudarshan, 
J. Math. Phys. 17, 821 (1976).

\bibitem{giovannetti2012}
V. Giovannetti, and G.M. Palma,
%Master Equations for correlated quantum channels,
Phys. Rev. Lett. 108, 040401 (2012).

\bibitem{ziman_deco}
M. Ziman, and V. Bu\v zek,
%All (qubit) decoherences: Complete characterization and physical implementation | pdf |
Phys. Rev. A 72, 022110 (2005).
%[quant-ph/0505040].
 
\bibitem{ziman_homo}
M. Ziman, P. \v Stelmachovi\v c, V. Bu\v zek, M. Hillery, V. Scarani, and N. Gisin,
%Dilluting quantum information: An analysis of information transfer in system-reservoir interactions | pdf |
Phys. Rev. A 65 , 042105 (2002).
%[quant-ph/0110164].

\bibitem{scarani_thermo}
V. Scarani, M. Ziman, P. \v Stelmachovi\v c, N. Gisin, V. Bu\v zek,
%Thermalizing Quantum Machines: Dissipation and Entanglement,
Phys. Rev. Lett. 88, 97905-1 (2002).

\bibitem{buzek_not}
V. Bu\v zek, M. Hilery, and R. Werner,
%Optimal manipulations wih qubits: Universal NOT gate,
Phys. Rev. A 60, R2626-R2629 (1999). 

\bibitem{martini} 
F. DeMartini, V. Bu\v{z}ek,  F. Sciarrino, and C.Sias,
%Experimental realization of the quantum universal NOT gate.''  
Nature  419, No. 6909, 815--819 (2002).

\bibitem{paulsen2003}
V.I.Paulsen, 
\emph{Completely Bounded Maps and Operator Algebras}, 
Cambridge Studies in Advanced Mathematics 78
(Cambridge University Press, Cambridge, 2003).

\bibitem{breuer2009_nonmarkovianity_measure}
H.P. Breuer, E.M. Laine, and J. Piilo,
%Measure for the Degree of Non-Markovian Behavior of Quantum Processes in Open Systems
Phys. Rev. Lett. 103, 210401 (2009). 
%[arXiv.org:0908.0238].

\bibitem{rivas2010}
A. Rivas, S.F. Huelga, and M.B. Plenio, 
%Entanglement and non-Markovianity of quantum evolutions 
Phys. Rev. Lett. 105 050403 (2010). 
%[arXiv.org:0911.4270].

\bibitem{usha_devi_2011}
A.R. Usha Devi, A.K. Rajagopal, and Sudha,
%Open-system quantum dynamics with correlated initial states, not completely positive maps and non-Markovianity,
Phys. Rev. A 83, 022109 (2011).

\end{thebibliography}
\end{document}